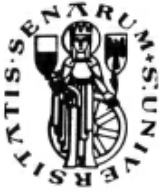

UNIVERSITÀ DEGLI STUDI DI SIENA

**Dipartimento di Economia Politica**

# Seats at the table: the network of the editorial boards in information and library science

Alberto Baccini [a] – Lucio Barabesi [b]

[a] Università di Siena - Dipartimento di Economia Politica, Via P.A. Mattioli, 10 I53100 SIENA
alberto.baccini@unisi.it tel. + 390577235233 fax +390577232661(Corresponding author)
[a] Università di Siena - Dipartimento di Economia Politica, Piazza San Francesco, 7 I53100 SIENA
barabesi@unisi.it tel. + 390577232796 fax +390577232661

ABSTRACT: The structural properties of the network generated by the editorial activities of the members of the boards of "Information Science & Library Science" journals are explored through network analysis techniques. The crossed presence of scholars on editorial boards, the phenomenon called interlocking editorship, is considered a proxy of the similarity of editorial policies. The evidences support the idea that this group of journals is better described as a set of only relatively connected subfields. In particular two main subfield are identified, consisting of research oriented journals devoted respectively to LIS and MIS. The links between these two subsets are weak. Around these two subsets there are a lot of (relatively) isolated professional journals or journals characterized more by their subject-matter content than by their focus on information flows. It is possible to suggest that this configuration of the network may be the consequence of the youthfulness of Information Science & Library Science, which has not permitted yet to reach a general consensus through scholars on research aims, methods and instruments.

# 1. Introduction

The domain of the present research is the academic community of information science and library science. This community is tentatively explored through the observation of the editorial activities of scholars engaged as members in the boards of editors of relevant scientific journals. The aim is to explore the structural properties of the network generated by the editorial activities of the members of the boards of these journals. While a lot of literature on sociology of science uses data on editorial boards for empirical research (e.g. Braun, 2004), starting at least from the seminal work of Merton and Zuckerman (1971); only recently these data have been explored with network analysis techniques (A. Baccini, 2009; Alberto Baccini & Barabesi, 2010; A. Baccini, Barabesi, & Marcheselli, 2009).

Traditionally, the main function of the editorial boards was to determine which articles were appropriate for publication. In the last two or three decades this function has changed: the spread of the anonymous referee process allows editorial boards to concentrate on selecting and evaluating referees (Hames, 2007; Powell, 2010). In any case, the role of editors can be considered of relevance in guiding research in a discipline, encouraging or suppressing various directions. No literature presents extensive discussions about the role of the board of editors for scientific journals (for a short overview see Alberto Baccini & Barabesi, 2010). The basic idea is that scholars can exercise influence on their scientific field by acting as the gatekeepers of the editorial policies of the journals (Braun & Diospatonyi, 2005; Braun, Diospatonyi, Zàdor, & Zsindely, 2007).

From a different point of view, scientific journals (and their publishers) are interested in assuring the presence of distinguished scholars in their boards. A cornerstone of the scientific ethos is that editorial board members should be selected based on their scholarly achievements (Bedeian, Van Fleet, & Hyman Iii, 2009). The competition between journals for scarce talented scholars results in partial overlapping of their editorial boards. If each member of the editorial board may influence in some measure the editorial policy of his/her journal, journals with overlapping boards may have partial overlapping editorial policies; or partially overlapping or complementary scopes. We will not be concerned with direct observations of the editorial policies adopted by the boards of journals, and of their contents –fields, subjects and methods covered. We will infer considerations about the similarity of editorial policies and consequently of journal contents by observing the crossed presence of scholars on editorial boards, a phenomenon called interlocking editorship (Alberto Baccini & Barabesi, 2010).

The scientific community of information and library science is represented as a network in which the vertices are journals and a link between a pair of journals is generated by the presence of a common editor on the board of both. Actually, this network is generated by a simple transformation of the so-called dual-mode or affiliation network. More precisely, a dual-mode network is one in which the vertices are divided into two sets (actors and events) and the affiliation connects the vertices from the two different sets only (de Nooy, Mrvar, & Batagelj, 2005; Wasserman & Faust, 1994, pp. 148-150). Dual-mode networks characterize some informetric phenomena: the author-paper links result in co-authorship/publication networks; the source-citation links result in reference-citation networks. In our case, the event of affiliation (being a member of the editorial board) connects a scholar to an information science journal. The duality specifically refers to the two alternative perspectives by which editors are linked by their affiliation to the same journal, and at the same time two journals are linked by the editors who are on their boards. Therefore, there are two different ways to view the affiliation network: as one of editors linked by journals (networks of co-membership), or as one of journals linked by editors (interlocking of events). It is possible to study the dual-mode network as a whole, or to transform the original dual-mode network into two single-mode networks focusing only on the analysis of the network of editors or of journals. Blaise Cronin (2009), calling attention to this exploratory approach for the information and library science community, has underlined the relevance of both perspectives. In this paper the focus is on the network of journals. By studying the structure of the information and library science journals network with the tools of network analysis, we can shed some light on the underlying processes according to which research is conducted by scholars. Our aims are (i) to establish which journals have a central position in the network and which a peripheral one; and (ii) to identify the groups, if any, in which the information and library science community break down.

## 2. The centre and periphery in the interlocking editorship network

The affiliation network database was constructed *ad hoc* for this paper. The journals considered are the 61 included in the category "Information Science & Library Science" of the 2008 edition of the *Journal of Citation Report Social Science Edition* managed by ISI-Thomson.[1] Other scholars adopt the same list of journals as representative (Bar-Ilan, 2010), but this choice may be considered controversial, excluding some research lines, e.g. information retrieval, that some scholars retain included in information and library science. Probably the best alternative strategy of journals selection consists in considering the list of (near) all the scientific journals relevant for the information science community. In other scientific communities such a list exists: for example, the more than a thousand journals considered in the *EconLit* (http://www.aeaweb.org/econlit/journal_list.php) database are considered the complete list of the relevant economic journals; analogously the *Philosopher Index* (http://philindex.org/) and *PubMed* (http://www.ncbi.nlm.nih.gov/pubmed) are considered containing all relevant scientific journals for, respectively, philosophy and medicine. Actually such a list does not exist for information and library science; the list compiled by Böll (2007) is not universally shared as the ones mentioned above in the respective scientific communities. This is the main reason conducting us to the choice of ISI-Thomson list.

The data on the members of the editorial boards of the 61 journals considered was directly obtained from the website of the journals. The data was collected in May 2010 considering the boards published on the websites of the journals in that period. Moreover, the database was managed by means of the package *Pajek* (Batagelj & Mrvar, 2006; de Nooy, et al., 2005).

There is no evidence regarding the roles of different kinds of editors in the editorial process (possibly apart from the role of editor-in-chief) and a single title such as managing editor may often entail very different roles for different journals. As a consequence a very broad notion of editor is adopted, covering all the individuals listed as editor, co-editor, member of the editorial board or of the advisory editorial board (Alberto Baccini & Barabesi, 2010; Braun & Diospatonyi, 2005; Hodgson & Rothman, 1999).

---

[1] Information Science & Library Science covers, according to the source considered, resources on a wide variety of topics, including bibliographic studies, cataloguing, categorization, database construction and maintenance, electronic libraries, information ethics, information processing and management, interlending, preservation, scientometrics, serials librarianship, and special libraries.

In this database, 2,003 seats were available on the editorial boards and they were occupied by 1,752 scholars. The average number of seats per journal turned out to be 32.8, while the average number of seats occupied by each scholar (*i.e.* the mean rate of participation) was 1.14. The number of lines linking the journals is 162, and the density of the interlocking directorship network (*i.e.* the ratio of the actual number of lines to the maximum possible number of lines in the network) is 0.087. This means that about 9% of the possible lines is present (Wasserman & Faust, 1994, pp. 314-317). These data depicts a network less connected than the one of statistical journals, having a similar dimension but higher density and mean participation rate (A. Baccini, et al., 2009).

The graph of the network is reported in Figure 1. The vertices in the graph are automatically placed by the package *Pajek* on the basis of the Kamada-Kawai algorithm that produces regularly spaced results for relatively small network (de Nooy, et al., 2005, pp. 16-17).

The degree distribution of the journals is contained in Table 1. The mean degree is 5.3 (while the median degree turns out to be 5) and the degree standard deviation is 4.66. All these value are lower than the corresponding value calculated for other disciplinary sectors, namely economics and statistics (Alberto Baccini & Barabesi, 2010; A. Baccini, et al., 2009).

**Figure 1 about here**
**Figure 1.** The information science journals network
(journals are labeled according to the legend of Table 2).

Ten journals are isolated from the network (*i.e.* they have zero degree). Quite a few of the isolates in Fig. 1 are either very librarianship- and practitioner-related (*Law Library Journal*, *Library Trends*, *Library Journal*, *Journal of the Medical Library Association*) or very specialized in nature (*International Journal of Geographical Information Science*; *Restaurator*). The isolation of *Library and Information Science* is due to insularity: it is in fact the journal of the Japanese Society for Library and Information Science, it is published in Japanese and its editorial board coincides with the editorial committee of the Society. Finally the idiosyncratic structure of the editorial boards determines the isolation of *Information Technology and Libraries* and *Learned Publishing*; the first has only one editor and the board of the second is composed by three members.

All the other journals are linked directly or indirectly; but in Figure I two subsets of journals may be impressionistically recognized. This first sight distinction reflects the categorization of journals proposed by Sugimoto, Pratt & Hauser (2008): in the upper part of the graph there are journals focused on management information science (MIS), in the lower part journals focused on library and information science (LIS). From this point of view, the so-called field as defined by ISI-Thomson database would be better described as a cluster of sub-fields, only partially interconnected. This first impression will be strengthened by a more formal analysis of the network structure.

**Table 1 about here**

**Table 1.** Degree frequency distribution of the Library & Information Science journals.

A main concern in network analysis is to distinguish between the centre and the periphery of the network. In our case, the problem is to distinguish between the journals which have a central position in the network and those in the periphery. As suggested by Wasserman and Faust (1994, pp. 187-192), three centrality measures for each journal in the network may be adopted. The simplest measure for the centrality of a journal is represented by its degree: indeed, the more ties a journal has to other journals, the more central its position in the network. For example, the *Journal of Documentation* is linked with 16 journals, while *Research Evaluation* is linked with solely one. Hence, the first is more central in the network than the second. In addition, the normalized degree of a journal is the ratio of its degree to the maximum possible degree (*i.e.* the number of journals minus 1). Thus, the *Journal of Documentation* is linked with about 27% of the other journals in the network, while *Research Evaluation* is linked with only 1.7%. Table 2 contains the degree and the normalized degree for the statistical journals considered. An overall measure of centralization in the network (based on marginal degrees) is given by so-called degree centralization (Wasserman & Faust, 1994, pp. 187-192). In this case, the index turns out to be 0.18, showing that the network of information science journals is less centralized than the other known disciplinary networks (economics and statistics).

The second centrality measure is given by closeness centrality, which is based on the distance between a journal and all the other journals. In the network analysis, the distance between two vertices is usually based on so-called geodesic distance. Geodesic is the shortest path between two vertices, while its length is the number of lines in the geodesic (Wasserman & Faust, 1994, pp. 187-192). Hence, the closeness centrality of a journal is the number of journals (linked to this journal by a path) divided by the sum of all the distances (between the journal and the linked journals). The basic idea is that a journal is central if its board can quickly interact with all the other boards. Journals occupying a central location with respect to closeness can be very effective in communicating information (sharing research, sharing papers, deciding editorial policies) to other journals. Table 2 contains the closeness centrality for information science journals. By focussing on the connected network of 51 journals, it is possible to compute the overall closeness centrality of journals (Wasserman & Faust, 1994, pp. 187-192). The overall closeness centrality is 0.32, showing in turn that this part of the network of information science journals is centralized in the same measure of other known journals network.

**Table 2 about here**

**Table 2.** Centrality measures and corresponding rankings of the information science journals

The third considered measure is the so-called betweeness centrality. The idea behind the index is that similar editorial aims between two non-adjacent journals might depend on other journals in the network, especially on those journals lying on the paths between the two. The other journals potentially might have some control over the interaction between two non-adjacent journals. Hence, a journal is more central in this respect if it is an important intermediary in links between other journals. From a formal perspective, the betweeness centrality of a journal is the proportion of all paths between pairs of other journals that include this journal. Table 2 contains the betweeness centrality of the journals. The journal with the highest betweeness centrality is *Information Society* which is in about 17% of the paths linking all other journals in the network. In fact, it is easy to see that this journal is the gatekeeper providing the links between the two subsets of LIS and MIS journals. More in general, the overall betweeness centralization of the network, that is the ratio of the variation in betweenes centrality scores to the maximum possible variation in a network of similar dimension (Wasserman & Faust, 1994, pp. 187-192) is 0.21, much higher than the corresponding measure in the network of the statistical journals (0.09). This result can be read as a consequence of the existence of relatively separated subfields connected through the editorial board of a minority of journals, permitting information flows between the subfields.

**3. Valued network analysis**

It is interesting to consider the strength of the relation between journals. The network of journals can be characterized as a valued network. More precisely, in a valued network the lines have a value indicating the strength of the tie linking two vertices (Wasserman & Faust, 1994, pp. 277-278). In our case the value of the line is the number of editors sitting on the board of the two journals linked by that line.

Table 3 shows the distribution of journals according to their line values: 55.6% of the links are generated by journals sharing only one editor and about 85% are generated by journals sharing three or less editors.

**Table 3 about here**

**Table 3.** Line multiplicity frequency distribution

In social network analysis it is usual to consider lines with higher value to be more important since they are less personal and more institutional (de Nooy, et al., 2005, p. 109). In the case of the journal network, the basic idea is very simple: the editorial proximity between two journals can be measured by observing the degree of overlap among their boards. Two journals with no common editors have no editorial relationship. Two journals with the same board share the same aim, *i.e.* the two journals have a common or, at least shared, editorial policy. Obviously, there are different degrees of integration between these two extreme cases. Actually, two journals sharing solely one member of their boards are less linked than two journals sharing two or more editors.

In information and library science there are not extreme cases of journals sharing all their editors, but there is a percentage of journals sharing more than three editors higher than in the other known scientific communities.

Starting from this basis it is possible to define cohesive subgroups, *i.e.* subsets of journals among which there are relatively strong ties. In a valued network a cohesive subgroup is a subset of vertices among which ties have a value higher than a given threshold. In our case, a cohesive subgroup of journals is a set of journals sharing a number of editors equal or higher than the threshold. In our interpretation, a cohesive subgroup of journals is a subgroup with a similar editorial policy, belonging to the same subfield of the discipline or sharing a common methodological approach. Following (de Nooy, et al., 2005, p. 109), cohesive subgroups are identified as weak components in *m*-slices, *i.e.* subsets for which the threshold value is at least *m*.

As previously remarked, the network of information and library science journals is relatively compact: with the exception of the ten isolated journals, it is possible to reach a given journal starting from any other journal. The search for cohesive subgroups confirms the presence of the two subsets impressionistically individuated in Figure 1, and the complete fragmentation of the others journals in groups mostly including solely one journal. Figure 2 represents the biggest component of the network containing a group of journals that can be considered as the LIS subfield. This component is identified as a weak component in 3-slices, that is the 13 journals in this subset of the network have at least 3 common editors. The density of this component is 0.24 indicating that a quarter of the possible links in the network are realized. The dimension of each vertex represents the betweeness centrality of the corresponding journal.

The centre of this component is represented by a complete subnetwork of four journals exclusively research-oriented and not geared to the interests of working professionals (librarians etc.): the *Journal of the American Society for Information Science and Technology*, the *Journal of Informetrics*, *Information Research* and the *Annual Review of Information Science and Technology*. This last title is a publication of the American Society for Information Science, it appears once each year and contains a total of, about, 12 commissioned chapters. Actually it shares the editor and five member of the board with the journal of the same society (JASIST); so it can be considered properly as a companion publication of JASIST. This is probably the reason for which it is the only journal of this complete subnetwork that does not control links toward other parts of LIS component, while through the others it is possible to reach all other parts. It is interesting to note that by dropping one of these four journals from the network, the structure of the network does not change. This point can be explained in reference to a sound editorial structure of the LIS subfield, in which the information flows do not depend from the role played by a single central journals, as happened for example in statistics where by dropping some journals the structure of the network collapses in isolated small disciplinary groups.

**Figure 2 about here**

**Figure 2.** The LIS weak component in 3-slices network
(the dimension of vertices is proportional to betweeness centrality).

The *Journal of Informetrics*, in turn, shares 16 board members with *Scientometrics*: this is the strongest link between two boards in the library and information science domain; it connects two leading journals in the explosive field (Van Noorden, 2010) of study dedicated to quantitative and bibliometric methods and applications.

The dimension of the vertices in the figures is proportional to the betweenness centrality in the general network; so it is easy to note in the Figure 2 the central role of JASIST, but also of other journals that despite their peripheral position in this component, control the links with the other parts of the network.

The second relevant component is drafted in Figure 3, it contains 9 journals and can be interpreted as the MIS subfield. This component is characterized by a high density of 0.472 indicating that about an half of the possible links between journals are realized. These links have also high value indicating that the numbers of common editors between journal is relatively high. This component is more strongly connected than the first one. *Information Society* is the journal with the highest betweeness centrality. In effect its (weak) link with *International Journal of Information Management* permits the flows of information between the two main components of the network.

It is worth to note that there are two other small component. The first contains two professional journals (*Online* (Wilton, Connecticut) and *E-Content*) mainly devoted to the applications of technology; and the second two professional journals for librarians *Library Quarterly* and *Library and Information Science Research*, and a more policy oriented journal (*Government Information Quarterly*).

**Figure 3 about here**

**Figure 3.** The MIS weak component in 3-slices network
(the dimension of vertices is proportional to betweeness centrality).

## 4. Conclusion

The exploratory analysis developed in this paper relies on the hypothesis that each editor possesses some power in the definition of the editorial policy of his/her journal. Consequently, if the same scholar sits on the board of two journals, those journals could have some common elements in their editorial policies. The proximity of the editorial policies of two scientific journals can be assessed by the number of common editors sitting on their boards. The degree of overlapping of the editorial boards of journals, called interlocking editorship, can be addressed with network analysis techniques.

For the Information Science & Library Science journals, the network generated by interlocking editorship seems to be not so compact as in other older and well established disciplines as economics and statistics. The Information Science & Library Science field as defined in ISI-Thomson is probably better described as a set of only relatively connected subfields. In particular two main components are identified, consisting of research oriented journals devoted respectively to LIS and MIS. The links between these two components are weak: a result completely coherent with the evidences drawn from cocitation analysis on the same fields (Sugimoto, et al., 2008). Around these components there are a lot of (relatively) isolated professional journals or journals characterized more by their subject-matter content than by their focus on information flows. At this stage of our knowledge it is only possible to conjecture that this configuration of the network may be the consequence of the relative youthfulness of Information Science & Library Science, which has not permitted yet to reach a general consensus through scholars on research aims, methods and instruments.

**Acknowledgements**

We would like to thank Blaise Cronin, whose comments on previous versions of this paper permit us a substantial improvement of our understanding of information and library science literature. Usual disclaimers apply.

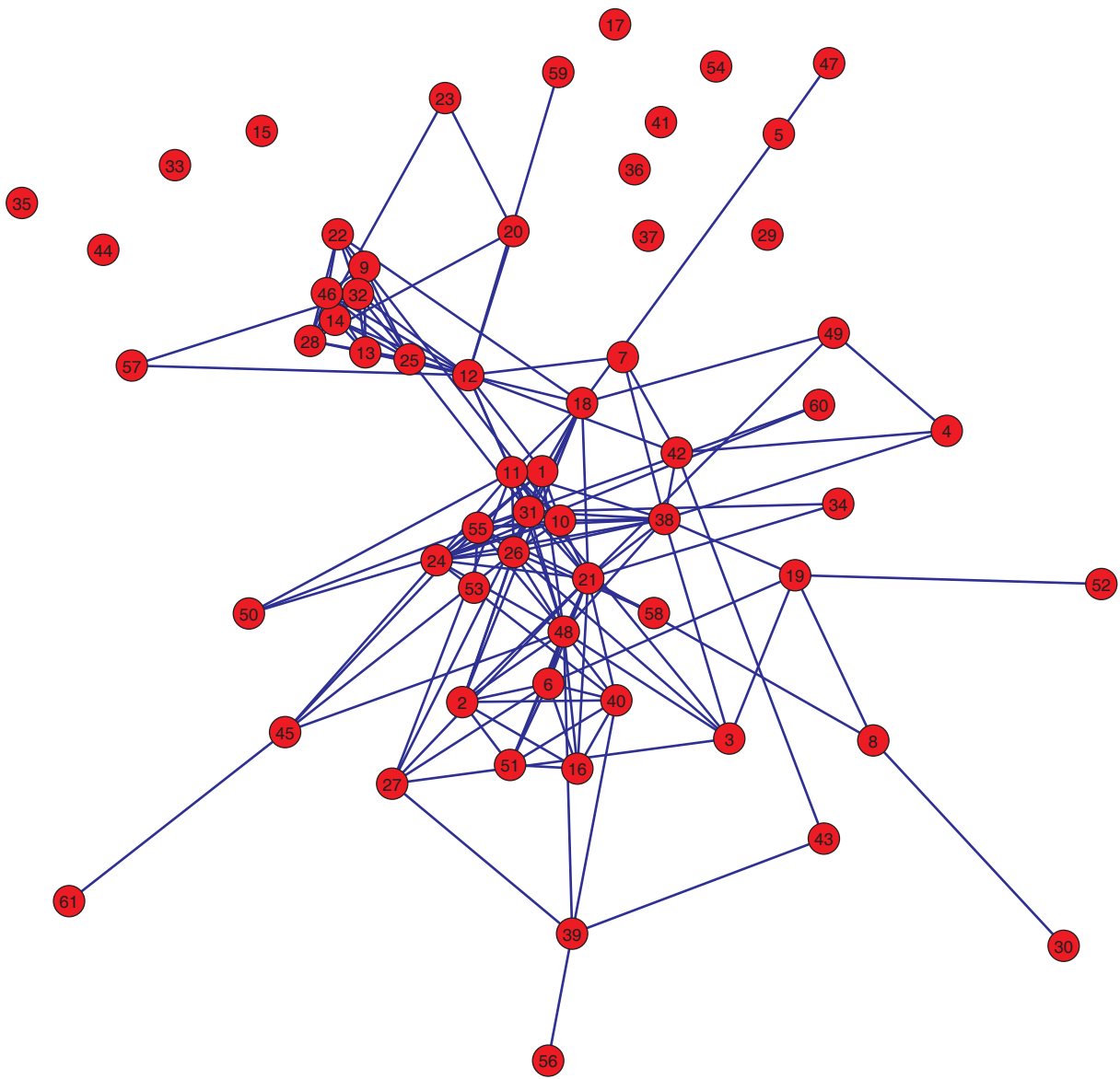

**Figure 1.** The information science journals network (journals are labeled according to the legend of Table 2).

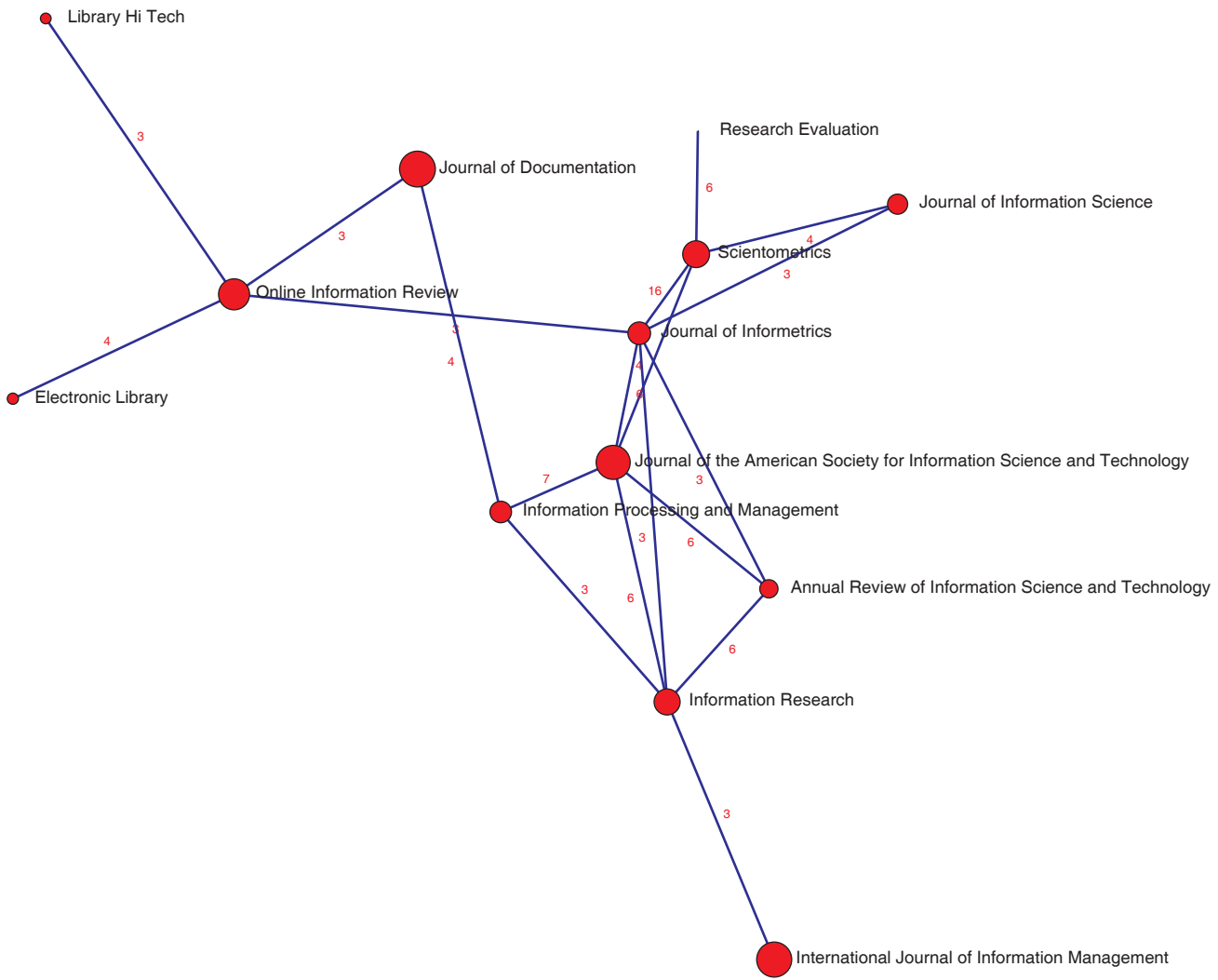

**Figure 2.** The LIS weak component in 3-slices network (the dimension of vertices is proportional to betweeness centrality).

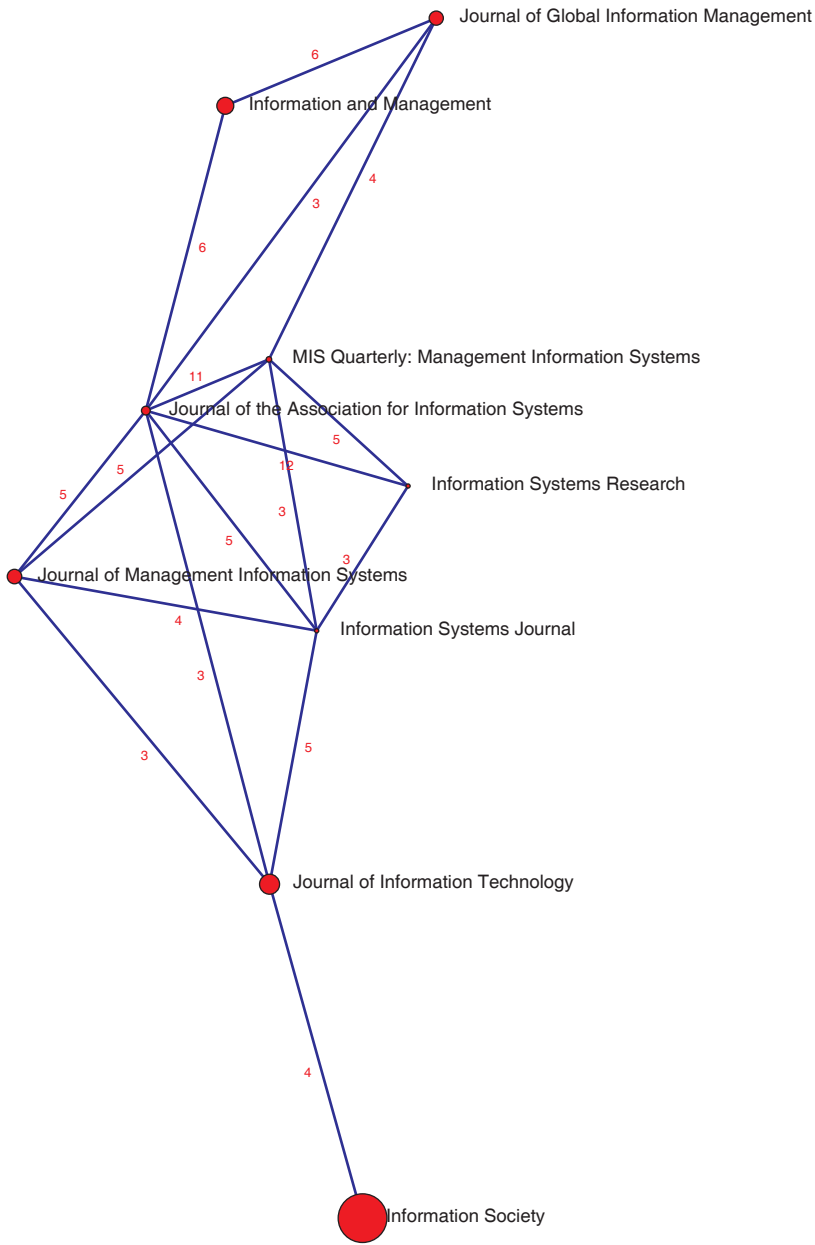

**Figure 3.** The MIS weak component in 3-slices network
(the dimension of vertices is proportional to betweeness centrality).

**Table 1.** Degree frequency distribution of the Library & Information Science journals.

| Degree | Freq | Freq% | CumFreq (%) |
|---|---|---|---|
| 0 | 10 | 0,164 | 0,164 |
| 1 | 7 | 0,115 | 0,279 |
| 2 | 7 | 0,115 | 0,393 |
| 3 | 6 | 0,098 | 0,492 |
| 5 | 4 | 0,066 | 0,557 |
| 6 | 5 | 0,082 | 0,639 |
| 7 | 2 | 0,033 | 0,672 |
| 8 | 4 | 0,066 | 0,738 |
| 9 | 4 | 0,066 | 0,803 |
| 10 | 3 | 0,049 | 0,852 |
| 12 | 2 | 0,033 | 0,885 |
| 13 | 3 | 0,049 | 0,934 |
| 14 | 2 | 0,033 | 0,967 |
| 16 | 2 | 0,033 | 1,000 |
|  | 61 |  |  |

**Table 2.** Centrality measures and corresponding rankings of the information science journals

| Label | Journal | Degree | Normalized Degree | Normalized Degree Rank | Closeness | Closeness Rank | Betweeness | Betweeness Rank | IF | IF rank |
|---|---|---|---|---|---|---|---|---|---|---|
| 1 | Annual Review of Information Science and Technology | 10 | 0,167 | 10 | 0,406 | 6 | 0,027 | 19 | 2,5 | 4 |
| 2 | Aslib Proceedings | 9 | 0,150 | 13 | 0,364 | 13 | 0,014 | 25 | 0,493 | 38 |
| 3 | Canadian Journal of Information and Library Science | 6 | 0,100 | 23 | 0,351 | 14 | 0,008 | 28 | 0 | 61 |
| 4 | College and Research Libraries | 3 | 0,050 | 32 | 0,307 | 29 | 0,002 | 33 | 0,781 | 31 |
| 5 | EContent | 2 | 0,033 | 38 | 0,281 | 37 | 0,028 | 17 | 0,271 | 55 |
| 6 | Electronic Library | 8 | 0,133 | 17 | 0,324 | 21 | 0,010 | 26 | 0,393 | 43 |
| 7 | Government Information Quarterly | 3 | 0,050 | 32 | 0,332 | 18 | 0,003 | 32 | 1,91 | 13 |
| 8 | Health information and libraries journal | 3 | 0,050 | 32 | 0,292 | 33 | 0,029 | 14 | 0,939 | 30 |
| 9 | Information and Management | 8 | 0,133 | 17 | 0,327 | 20 | 0,020 | 20 | 2,358 | 6 |
| 10 | Information Processing and Management | 10 | 0,167 | 10 | 0,402 | 7 | 0,038 | 11 | 1,852 | 15 |
| 11 | Information Research | 13 | 0,217 | 5 | 0,435 | 2 | 0,055 | 8 | 1 | 28 |
| 12 | Information Society | 14 | 0,233 | 3 | 0,394 | 10 | 0,166 | 1 | 1,042 | 27 |
| 13 | Information Systems Journal | 7 | 0,117 | 21 | 0,290 | 35 | 0,001 | 35 | 2,375 | 5 |
| 14 | Information Systems Research | 7 | 0,117 | 21 | 0,290 | 35 | 0,001 | 35 | 2,261 | 9 |
| 15 | Information Technology and Libraries | 0 | 0,000 | 52 | 0,000 | 52 | 0,000 | 38 | 0,703 | 33 |
| 16 | Interlending and Document Supply | 6 | 0,100 | 23 | 0,317 | 23 | 0,000 | 38 | 1,596 | 20 |
| 17 | International Journal of Geographical Information Science | 0 | 0,000 | 52 | 0,000 | 52 | 0,000 | 38 | 1,043 | 26 |
| 18 | International Journal of Information Management | 10 | 0,167 | 10 | 0,410 | 5 | 0,099 | 3 | 0,484 | 39 |
| 19 | Journal of Academic Librarianship | 5 | 0,083 | 28 | 0,301 | 30 | 0,033 | 12 | 0,667 | 35 |
| 20 | Journal of Computer-Mediated Communication | 3 | 0,050 | 32 | 0,281 | 37 | 0,016 | 21 | 3,428 | 2 |
| 21 | Journal of Documentation | 16 | 0,267 | 1 | 0,414 | 3 | 0,104 | 2 | 1,954 | 12 |
| 22 | Journal of Global Information Management | 6 | 0,100 | 23 | 0,312 | 26 | 0,014 | 24 | 1,836 | 16 |
| 23 | Journal of Health Communication | 2 | 0,033 | 38 | 0,221 | 48 | 0,000 | 38 | 1,901 | 14 |
| 24 | Journal of Information Science | 12 | 0,200 | 8 | 0,394 | 10 | 0,033 | 13 | 1,712 | 17 |
| 25 | Journal of Information Technology | 9 | 0,150 | 13 | 0,351 | 14 | 0,028 | 16 | 1,387 | 21 |
| 26 | Journal of Informetrics | 13 | 0,217 | 5 | 0,402 | 7 | 0,041 | 9 | 2,057 | 10 |
| 27 | Journal of Librarianship and Information Science | 6 | 0,100 | 23 | 0,348 | 16 | 0,016 | 22 | 1,648 | 19 |
| 28 | Journal of Management Information Systems | 9 | 0,150 | 13 | 0,279 | 40 | 0,015 | 23 | 1,966 | 11 |
| 29 | Journal of Scholarly Publishing | 0 | 0,000 | 52 | 0,000 | 52 | 0,000 | 38 | 2,531 | 3 |
| 30 | Journal of the American Medical Informatics Association | 1 | 0,017 | 45 | 0,218 | 50 | 0,000 | 38 | 0,562 | 36 |
| 31 | Journal of the American Society for Information Science and Technology | 14 | 0,233 | 3 | 0,449 | 1 | 0,094 | 4 | 2,358 | 6 |
| 32 | Journal of the Association for Information Systems | 9 | 0,150 | 13 | 0,294 | 31 | 0,005 | 29 | 1,669 | 18 |
| 33 | Journal of the Medical Library Association : JMLA | 0 | 0,000 | 52 | 0,000 | 52 | 0,000 | 38 | 0,455 | 40 |
| 34 | Knowledge Organization | 2 | 0,033 | 38 | 0,314 | 25 | 0,000 | 38 | 0,429 | 41 |
| 35 | Law Library Journal | 0 | 0,000 | 52 | 0,000 | 52 | 0,000 | 38 | 0,296 | 53 |
| 36 | Learned Publishing | 0 | 0,000 | 52 | 0,000 | 52 | 0,000 | 38 | 0,559 | 37 |
| 37 | Library and Information Science | 0 | 0,000 | 52 | 0,000 | 52 | 0,000 | 38 | 0,364 | 46 |
| 38 | Library and Information Science Research | 13 | 0,217 | 5 | 0,402 | 7 | 0,079 | 5 | 0,344 | 50 |
| 39 | Library Collections, Acquisition and Technical Services | 5 | 0,083 | 28 | 0,294 | 31 | 0,038 | 10 | 0,091 | 59 |
| 40 | Library Hi Tech | 8 | 0,133 | 17 | 0,329 | 19 | 0,009 | 27 | 1,226 | 23 |
| 41 | Library Journal | 0 | 0,000 | 52 | 0,000 | 52 | 0,000 | 38 | 0,388 | 44 |
| 42 | Library Quarterly | 5 | 0,083 | 28 | 0,345 | 17 | 0,028 | 15 | 0,364 | 46 |

| | | | | | | | | | |
|---|---|---:|---:|---:|---:|---:|---:|---:|---:|
| 43 | Library Resources and Technical Services | 2 | 0,033 | 38 | 0,270 | 43 | 0,004 | 31 | 0,698 | 34 |
| 44 | Library Trends | 0 | 0,000 | 52 | 0,000 | 52 | 0,000 | 38 | 0,239 | 56 |
| 45 | Libri | 5 | 0,083 | 28 | 0,310 | 28 | 0,028 | 17 | 0,156 | 58 |
| 46 | MIS Quarterly: Management Information Systems | 8 | 0,133 | 17 | 0,292 | 33 | 0,002 | 34 | 5,183 | 1 |
| 47 | Online (Wilton, Connecticut) | 1 | 0,017 | 45 | 0,211 | 51 | 0,000 | 38 | 0,352 | 49 |
| 48 | Online Information Review | 16 | 0,267 | 1 | 0,414 | 3 | 0,078 | 6 | 1,103 | 25 |
| 49 | portal: Libraries and the Academy | 3 | 0,050 | 32 | 0,324 | 21 | 0,004 | 30 | 1,146 | 24 |
| 50 | Profesional de la Informacion | 3 | 0,050 | 32 | 0,312 | 26 | 0,000 | 37 | 0,4 | 42 |
| 51 | Program | 6 | 0,100 | 23 | 0,317 | 23 | 0,000 | 38 | 0,286 | 54 |
| 52 | Reference and User Services Quarterly | 1 | 0,017 | 45 | 0,222 | 47 | 0,000 | 38 | 0,339 | 52 |
| 53 | Research Evaluation | 1 | 0,017 | 45 | 0,265 | 45 | 0,000 | 38 | 1 | 28 |
| 54 | Restaurator | 0 | 0,000 | 52 | 0,000 | 52 | 0,000 | 38 | 0,172 | 57 |
| 55 | Scientometrics | 12 | 0,200 | 8 | 0,384 | 12 | 0,058 | 7 | 0,353 | 48 |
| 56 | Serials Review | 1 | 0,017 | 45 | 0,219 | 49 | 0,000 | 38 | 2,328 | 8 |
| 57 | Social Science Computer Review | 2 | 0,033 | 38 | 0,277 | 41 | 0,000 | 38 | 0,383 | 45 |
| 58 | Social Science Information | 2 | 0,033 | 38 | 0,281 | 37 | 0,000 | 38 | 0,714 | 32 |
| 59 | Telecommunications Policy | 1 | 0,017 | 45 | 0,270 | 43 | 0,000 | 38 | 0,341 | 51 |
| 60 | The Scientist | 2 | 0,033 | 38 | 0,277 | 41 | 0,000 | 38 | 1,244 | 22 |
| 61 | Zeitschrift für Bibliothekswesen und Bibliographie | 1 | 0,017 | 45 | 0,227 | 46 | 0,000 | 38 | 0,019 | 60 |

**Table 3.** Line multiplicity frequency distribution.

| Line value | Freq | Freq (%) |
|---|---|---|
| 1 | 90 | 55.6 |
| 2 | 33 | 20.4 |
| 3 | 15 | 9.3 |
| 4 | 8 | 4.9 |
| 5 | 5 | 3.1 |
| 6 | 7 | 4.3 |
| 7 | 1 | 0.6 |
| 8 | 0 | 0.0 |
| 9 | 0 | 0.0 |
| 10 | 0 | 0.0 |
| 11 | 1 | 0.6 |
| 12 | 1 | 0.6 |
| 13 | 0 | 0.0 |
| 14 | 0 | 0.0 |
| 15 | 0 | 0.0 |
| 16 | 1 | 0.6 |